\documentclass[dvips]{article}
\newcommand{\doublespacing}{\let\CS=\@currsize\renewcommand{\baselinesstrech}
{2.0}\tiny\CS}
\begin{document}

\textwidth 16cm
\newcommand{\bd}{\begin{document}}
\newcommand{\ed}{\end{document}}
\newcommand{\bc}{\begin{center}}
\newcommand{\ec}{\end{center}}
\newcommand{\bfr}{\begin{flushright}}
\newcommand{\efr}{\end{flushright}}
\newcommand{\lt}{\left}
\newcommand{\rt}{\right}
\newcommand{\vs}{\vspace}
\newcommand{\hs}{\hspace}
\newcommand{\beq}{\begin{equation}}
\newcommand{\eeq}{\end{equation}}
\newcommand{\lb}{\linebreak}
\newcommand{\pb}{\pagebreak}
\newcommand{\mb}{\makebox}
\newcommand{\fb}{\framebox}
\newcommand{\mc}{\multicolumn}
\newcommand{\ben}{\begin{enumerate}}
\newcommand{\een}{\end{enumerate}}
\newcommand{\bit}{\begin{itemize}}
\newcommand{\eit}{\end{itemize}}
\newcommand{\ol}{\overline}
\newcommand{\un}{\underline}
\newcommand{\lefq}{\lefteqn}
\newcommand{\ba}{\begin{array}}
\newcommand{\ea}{\end{array}}
\newcommand{\beqa}{\begin{eqnarray}}
\newcommand{\eeqa}{\end{eqnarray}}
\newcommand{\beqas}{\begin{eqnarray*}}
\newcommand{\eeqas}{\end{eqnarray*}}
\newcommand{\bfg}{\begin{figure}}
\newcommand{\efg}{\end{figure}}
\newcommand{\bds}{\begin{displaymath}}
\newcommand{\eds}{\end{displaymath}}
\newcommand{\btb}{\begin{tabbing}}
\newcommand{\etb}{\end{tabbing}}
\newcommand{\para}{\parallel}
\newcommand{\pad}{\partial}
\newcommand{\nn}{\nonumber}
\newcommand{\la}{\leftarrow}
\newcommand{\ra}{\rightarrow}
\newcommand{\lgla}{\longleftarrow}
\newcommand{\lgra}{\longrightarrow}
\newcommand{\La}{\Leftarrow}\newcommand{\Ra}{\Rightarrow}
\newcommand{\Lra}{\Leftrightarrow}
\newcommand{\Lgla}{\Longleftarrow}
\newcommand{\Lgra}{\Longrightarrow}
\newcommand{\bm}{\boldmath}
\newcommand{\lan}{\langle}
\newcommand{\ran}{\rangle}
\renewcommand{\a}{\alpha}
\renewcommand{\b}{\beta}
\newcommand{\g}{\gamma}
\newcommand{\G}{\Gamma}
\renewcommand{\d}{\delta}
\newcommand{\eps}{\epsilon}
\newcommand{\Th}{\Theta}
\newcommand{\s}{\sigma}
\newcommand{\lam}{\lambda}
\newcommand{\D}{\Delta}
\newcommand{\vare}{\varepsilon}
\newcommand{\pr}{\prime}
\newcommand{\ro}{\rho}
\newcommand{\nab}{\nabla}
\newcommand{\m}{\mu}
\newcommand{\n}{\nu}
\newcommand{\Sg}{\Sigma}
\newcommand{\p}{\pi}
\newcommand{\R}{I\!\!R}
\newcommand{\om}{\omega}
\newcommand{\Om}{\Omega}
\newcommand{\ze}{\zeta}
\newcommand{\vart}{\vartheta}
\newcommand{\tri}{\triangle}
\newcommand{\f}{\frac}
\newcommand{\iny}{\infty}
\newcommand{\pro}{\propto}
\pb
\bc {\huge A class of exactly solvable Schr\"odinger equation with moving boundary condition} \ec

\vs{1cm}

\bc
{\it T.K. Jana{\footnote {e-mail : tapas$_{-}$r@isical.ac.in} and P. Roy{{\footnote{e-mail : pinaki@isical.ac.in}}\\
Physics \& Applied Mathematics Unit \\
Indian Statistical Institute \\
Kolkata - 700 108, India.}}} \ec
\vs{4.5cm}

\bc {\large {\un{abstract}}} \ec 
Using first and second order supersymmetry formalism we obtain a class of exactly solvable potentials subject to moving boundary condition.\\

Keywords: moving boundary; supersymmetry.\\
PACS numbers: 03.65.-w,03.65.Ca
\pb
\section{Introduction}  Time dependent Schr\"odinger equations appear in many areas of quantum mechanics. Among the various time dependent problems, those subjected to moving boundary conditions are very interesting. Such a system was first considered by Fermi \cite{fermi} in connection with the study of cosmic radiation. Subsequently quantised moving boundary problems have been studied by a number of authors \cite{rice,munier,makowski1,makowski2,makowski3,dodonov,li,ho,ho1,yuce}. However, in most cases exact solutions have been obtained for a particle in a constant or harmonic oscillator potential subjected to moving boundary condition. In view of this it is of some interest to obtain other exactly solvable potentials subject to moving boundary conditions. 

A standard approach to the moving boundary problems is to transform them to problems with fixed boundary. In case the fixed boundary problem is exactly solvable then the original problem also becomes exactly solvable.  However, there are not many time independent potentials which are solvable over a fixed segment of the real line. On the other hand it is well known that the class of exactly solvable potentials can be enlarged using, for example, supersymmetric (SUSY) or Darboux formalism \cite{cooper,junker}. In the case of time dependent problems it is possible to enlarge the class of exactly solvable potentials by directly applying the time dependent Darboux transformation \cite{arrigo,sam,finkel}. However here we plan to follow a somewhat simpler procedure. More precisely we shall first apply the first (and second order) order time independent Darboux transformation and subsequently use the separation of variable technique to obtain new time dependent potentials. In particular we shall start with the square well potential with a moving right boundary and construct exactly solvable moving boundary potentials using first order as well as higher order SUSY formalism. The organisation of the paper is as follows: in section 2, we outline separation of variable approach to moving boundary problems; in section 3, we apply SUSY formalism to construct new exactly solvable moving boundary problems and finally section 4 is devoted to a conclusion.

\section{Separation of variables approach to Schr\"odinger equation with a moving boundary.}

The Schr\"odinger equation
\beq
\left[-\frac{\partial^2}{\partial x^2}+V(x,t)\right]\psi(x,t)=i\frac{\partial \psi(x,t)}{\partial t} \label{sch1} 
\eeq
endowed with the boundary condition
\beq
\psi(0,t)=0,~~~~~~~~~\psi(L(t),t)=0 \label{mvc} 
\eeq  
describes the moving boundary problem in quantum mechanics. Here $L(t)$ defines the expansion of the boundary.

We shall now use the separation of variable technique \cite{spector,spector1} to transform Eqn.(\ref{sch1}) to a problem with fixed boundary. To do this let us transform the variable $x\longmapsto q$ as \beq q=\frac{x}{L(t)}~~,~~0\leq q \leq 1 \label{tr1}
\eeq 

We now consider the potential to be of the form 
\beq
V(q,t)= g(t)\tilde{V}(q) + U(q,t) + g_{0}(t) \label{vq}
\eeq
Let us now transform the wave function as
\beq
\psi(q,t)\rightarrow e^{\phi(q,t)}\chi(q,t) \label{chiq}
\eeq
Using the transformations (\ref{tr1}) and (\ref{chiq}), Eqn.(\ref{sch1}) becomes
\beq
\ba{l}
-\frac{\partial^2\chi}{\partial q^2}+\frac{\partial\chi}{\partial q}\left[-2\frac{\partial\phi}{\partial q}+iq\dot{L}(t)L(t)\right] +\chi\left[-\left\{\frac{\partial^2\phi}{\partial q^2}+(\frac{\partial\phi}{\partial q})^2\right\}+i\frac{\partial\phi}{\partial q}q\dot{L}(t)L(t)-iL^2(t)\frac{\partial\phi}{\partial t}\right]\\\\
+\chi L^2(t)\left[g(t)\tilde{V}(q)+U(q,t)\right]=i\frac{\partial \chi}{\partial t}L^2(t)-L^2(t)g_0(t)\chi \label{sr}\ea\eeq 

For separability, the coefficient of $\frac{\partial\chi}{\partial q}$ should be a function of q. Thus we get
\beq
\phi(q,t)=a(t)\frac{q^2}{2}+b(q)+c(t) \label{phiqt}\eeq
where $a(t)=\frac{i}{2}\dot{L}(t)L(t)$. Let us choose $b(q)=0$ and \beq c(t)=-i\int^{t}_{0}g_{0}(s)ds -\frac{1}{2}log L(t)\label{ct}\eeq

On using (\ref{phiqt}) and (\ref{ct}) in Eqn.(\ref{sr}) we get
\beq
-\frac{\partial^2\chi}{\partial q^2}+\left[\frac{1}{4}L^3(t)\ddot{L}(t)q^2 +U(q,t)L^2(t) +g(t)\tilde{V}(q)L^2(t)\right]\chi=i\frac{\partial \chi}{\partial t}L^2(t) \label{sch3}\eeq
Now writing \beq \chi(q,t)=Q(q)T(t) \label{sep1}\eeq
we obtain from (\ref{sch3}) 
\beq
-\frac{Q^{''}}{Q}+\frac{1}{4}L^3(t)\ddot{L}(t)q^2 +L^2(t)g(t)\tilde{V}(q) +L^2(t)U(q,t)=i\frac{\dot{T}}{T}L^2(t)\label{schsep}\eeq
where primes represent derivative with respect to q. Now two cases may arise depending on the choice of $U(q,t)$:
\vspace{.2cm}
 
\un{Case I}: Let us consider $U(q,t)=0 $. In this case Eqn.(\ref{schsep}) separates if
 
\begin{enumerate}
\item \beq \hspace{-25em} g(t)=\frac{1}{L^2(t)}\label{gt}\eeq
\item \beq \hspace{.25em} L^3(t)\ddot{L}(t)=constant=c_1~~~i.e~ L(t)=\sqrt{\lambda t^2+\mu t+\nu},~~ where~~ c_1=\lambda\nu-\f{\mu^2}{4}. \label{l}\eeq
\end{enumerate}
It may be noted that in this case the boundary is moving with non uniform velocity.
Now from Eqn.(\ref{schsep}) we get 
\beq
-\frac{d^2Q}{dq^2}+v_1(q)Q=\epsilon Q \label{indeq}\eeq
where $v_1(q)=\tilde{V}(q)+\frac{c_1}{4}q^2$ and $\epsilon$ is the separation constant.\\
and
\beq
\frac{\dot{T}}{T}=-\frac{i\epsilon}{L^2(t)} \label{timeq}\eeq
i.e \beq T(t)=e^{-i\epsilon\tau(t)},~~ where~~ \tau(t)=\int^{t}_{0}\frac{1}{L^2(s)}ds \label{t}\eeq
\vspace{.2cm}

\un{Case II}: We now take \beq U(q,t)=-\frac{1}{4}L(t)\ddot{L}(t)q^2\label{uq}\eeq
Note that unlike in case I, the moving boundary is not subject to any particular law and may be chosen according to a particular situation. Then using (\ref{uq}) in (\ref{schsep}) it can be shown that the condition of separability requires $g(t)=\frac{1}{L^2(t)}$.
Therefore from Eqn.(\ref{schsep}) we again obtain two equations, one of which is
\beq 
-\frac{d^2Q}{dq^2}+\tilde{V}(q)Q=\epsilon Q \label{indeq1}\eeq
and other is identical with Eqn.(\ref{timeq}).\\
Thus in the first case if $v_1(q)$ is solvable potential then the original time dependent potential is also solvable. In the second case the original time dependent problem is solvable if $\tilde{V}(q)$ is a solvable potential. 
However in the second case there is a larger family of separable (consequently a larger family of exactly solvable) models as  $U(q,t)\neq 0$. In either case the wavefunctions are of the form
\beq
\psi(x,t) = \sum_n c_ne^{\phi(x,t)}Q_n(x,t)T_n(t)
\eeq

\section{Construction of potentials via SUSY} 

Here we shall construct some potentials appropriate for moving boundary problems using SUSY formalism \cite{cooper,junker}. In order to have a fairly general class of potentials we consider $U(q,t)\neq 0$ and for simplicity we choose $g_0(t)=0$.

As mentioned in the last section it is now necessary to choose $\tilde{V}$ and we consider a square well potential of unit length \cite{cooper,junker}
\beq
{\tilde V}(q) = -\pi^2 ~~~,~~~0 \leq q \leq 1\\
\label{vqexm}\eeq
The energy spectrum and eigenfunctions of this potential are well known and are given by \cite{cooper,junker}
\beq
\epsilon_{n}=n(n+2)\pi^2 ~~~~and~~~Q_{n}(q)\propto sin[(n+1)\pi q]
\label{sqew}\eeq
Then from Eqn.(\ref{sep1}) it follows that
\beq
\chi_{n}(q,t)=Q_{n}(q)T(t)\propto~sin[(n+1) \pi q]e^{-i\epsilon_{n}\tau(t)}\eeq
where $\tau(t)$ is given by Eqn.(\ref{t}).
Then from (\ref{vq}) the time dependent potential is found to be
\beq
V(x,t)=-\frac{\pi^2}{L^2(t)}-\frac{\ddot{L}(t)}{4L(t)}x^2~~,~~ 0\leq x\leq L(t)\label{evxt1}\eeq

The wavefunctions can be found from (\ref{chiq}) and are given by
\beq
\psi_{n}(x,t)\propto \frac{1}{\sqrt{L(t)}}e^{\frac{i}{4}\frac{\dot{L}(t)}{L(t)}x^2 -i\epsilon_{n}\tau(t)}sin[\frac{(n+1) \pi x}{L(t)}] \label{ewxt1}\eeq
where in obtaining (\ref{ewxt1}) we have used $\phi(q,t)=\frac{i}{4}\dot{L}(t) L(t) q^2 -\frac{1}{2} log L(t)$.
Thus under the moving boundary condition (\ref{mvc}) the time dependent Schr\"odinger equation (\ref{sch1}) is exactly solvable with the potential (\ref{evxt1}). 

\subsection {New potential via first order SUSY}
In first order SUSY formalism \cite{cooper,junker} two Hamiltonians $H_{0,1}=A^{\mp}A^{\pm}$ are intertwined through two first order differential operators of the form $A^{\pm} = \pm\f{d}{dx}+w(x)$, $w(x)$ being the superpotential. The two potentials $V_{\pm}=(w^2\pm w')$ are isospectral except perhaps the ground state. To cast the potential (\ref{vqexm}) into the SUSY form, we consider the superpotential to be of the form

\beq
w(q)=-\pi cot(\pi q)\label{ewq1}\eeq
Then the partner potentials are of the form 
\beq
V_{-}(q)=w^2(q)-w'(q)=- \pi^2\label{evm1}\eeq   
\beq
V_{+}(q)=w^2(q)+w'(q)= \pi^2\left[2~cosec^{2}(\pi q)-1\right]  \label{evp1}\eeq   
It is seen that $V_{-}(q)$ is a square well of unit length as in Eqn.(\ref{vqexm}). Also the wavefunctions of $H_{0,1}$ are related to each other because of SUSY and using (\ref{sqew}) 
the wavefunctions of $V_{+}(q)$ are found to be
\beq
Q_{n+}(q)\propto \left[\frac{d}{dq}+w(q)\right]Q_{(n+1)-}(q)
\eeq 
From (\ref{chiq}) and (\ref{sep1}) the wavefunctions are found to be 
\beq
\psi_{n+}(x,t)\propto \frac{\pi}{\sqrt{L(t)}}e^{\frac{i}{4}\frac{\dot{L}(t)}{L(t)}x^2 -i\epsilon_{n+}\tau(t)}\left[(n+2) cos(\frac{(n+2)\pi x}{L(t)})- cot(\frac{\pi x}{L(t)})sin(\frac{(n+2)\pi x}{L(t)})\right]
\eeq
where $\epsilon_{n+}=\epsilon_{(n+1)-}=(n+1)(n+3)$. 
\vspace{.2cm}

Thus the time dependent potential is given by
\beq
V_{+}(x,t)=\frac{\pi^2}{L^2(t)}\left[2~cosec^{2}(\frac{\pi x}{L(t)})-1\right]-\frac{\ddot{L}(t)}{4L(t)}x^2~~,~~ 0\leq x\leq L(t)\label{evxtp1}
\eeq
This is the new potential which solves exactly the Schr\"odinger equation (\ref{sch1}) under the moving boundary condition (\ref{mvc}).

\subsection{New potentials via second order SUSY}
In contrast to the previous case, in second order SUSY formalism the intertwining operators are second order differential operators of the form \cite{fernandez,sam1,sam2}
\beq
\ba{lcl}
D &=& {\displaystyle \f{d^2}{dq^2} + \beta(q)\f{d}{dq} + \gamma(\beta)}\\\\
\beta(q) &=& \displaystyle \f{(\epsilon^{0}_{j+1}-\epsilon^{0}_j)Q_{j}^{0}(q)Q^{0}_{j+1}(q)}{ W_{j,j+1}(q)}\\\\
\gamma(\beta) &=& {\displaystyle -\f{\beta''}{2\beta} + \left(\f{\beta'}{2\beta}\right)^2 +
\f{\beta'}{2} + \f{\beta^2}{4} - \left(\f{\epsilon^{0}_{j+1}-\epsilon^{0}_j}{2\beta}\right)^2}\\
\ea \label{D} \eeq 
where $Q^{0}_j$  are eigenfunctions of
$H_0$ (the Hamiltonion corresponding to the exactly solvable potential $V_{0}$) corresponding to the eigenvalues $\epsilon^{0}_j$ 
and $W_{j,j+1} = (Q^{0}_{j} Q^{0'}_{j+1}-Q^{0'}_{j} Q^{0}_{j+1})$ is the associated
Wronskian. Then the isospectral partner potential $V_2(q)$ of $V_{0}(q)$ obtained
via second order SUSY formalism is given by \cite{fernandez} \beq V_2(q) = V_0(q) -
2\f{d^2}{dq^2}\log W_{j,j+1}(q)\label{v2} \eeq The wave functions
$Q^{0}_j(q)$ and $Q^{2}_{j}(q)$ corresponding to $H_0$ and $H_2$
are connected by \beq Q^{2}_{k}(q) = \displaystyle{D Q^{0}_{k} (q) = \f{1}{W_{j,j+1}(q)}}\left|\ba{ccc} Q^{0}_{j} & Q^{0}_{j+1} & Q^{0}_{k}\\ Q^{0'}_{j} & Q^{0'}_{j+1} & Q^{0'}_{k}\\ Q^{0''}_{j} & Q^{0''}_{j+1} & Q^{0''}_{k}\\ \ea\right|,~~ j,j+1\neq k\label{phi} \eeq  The eigenfunctions
obtained from $Q^{0}_{j}$ and $Q^{0}_{j+1}$ are given by \beq f(q)\propto
\f{Q^{0}_{j}}{W_{j,j+1}(q)}~~,~~g(q)\propto
\f{Q^{0}_{j+1}}{W_{j,j+1}(q)} \label{phiij} \eeq  Here the
eigenfunctions $f(q),g(q)$ in (\ref{phiij}) are not acceptable
because they cannot be normalized. 

Therefore by using different consecutive levels of the exactly solvable potential $V_{0}(q)$ we can construct different $V_{2}(q)$. Then the exactly solvable time dependent potentials $V_{0}(x,t)$ and $V_{2}(x,t)$ can be obtained from (\ref{vq}). The wavefunctions of the corresponding time dependent potentials are given by
\beq
\psi_{n}^{0,2}(x,t)=e^{\phi(\frac{x}{L(t)},t)}Q_{n}^{0,2}(\frac{x}{L(t)})T_n(t)~~,(n\neq j, j+1)\label{v2wxt}\eeq
Here we can generate a large class of exactly solvable time dependent potentials as we can use any two consecutive levels. 

To obtain new potentials, we take the starting potential $V_0(q)$ to be the same as ${\tilde V}(q)$ in (\ref{vqexm}). The corresponding time dependent counterpart of $V_{0}(q)$ is $V_{0}(x,t)$ and states $\psi^{0}_{n}(x,t)$ are same as in Eqns.(\ref{evxt1}) and (\ref{ewxt1}) respectively.
\vspace{.25cm} 

\un{\bf{Construction of $V_{2}(q)$ for $j=0$}}:
\vspace{.2cm}

From (\ref{sqew}) we find
\beq
Q_{0}^{0}=sin(\pi q)~~and~~~Q_{1}^{0}=sin(2\pi q)\label{q01}\eeq
so that using (\ref{v2}) and (\ref{phi}) we get 
\beq
V_{2}(q)=\pi^2 (6~cosec^{2}(\pi q)-1)\label{v201}\eeq 
\beq
Q_{n}^{2}(q) = \pi^2 sin[(n+1)\pi q]\left\{3~cosec^{2}(\pi q)-(n^2+2n+3)\right\}
-3(n+1)\pi^2 cot(\pi q)cos[(n+1)\pi q]\label{v2qw01}\eeq
Therefore in this case the exactly solvable time dependent potential is
\beq
V_{2}(x,t)=\frac{\pi^2}{L^2(t)}\left[6~cosec^{2}(\frac{\pi x}{L(t)})-1\right]-\frac{\ddot{L}(t)}{4L(t)}x^2~~,~~ 0\leq x\leq L(t)\label{v201xt}
\eeq
and the corresponding wavefunctions are found to be (using eq.(\ref{v2wxt}) and (\ref{v2qw01}))
\beq
\ba{l}
\psi_{n}^{2}(x,t)\propto \frac{\pi^2}{\sqrt{L(t)}}e^{\frac{i}{4}\frac{\dot{L}(t)}{L(t)}x^2 -i\epsilon_{n2}\tau(t)}\left(sin[\frac{(n+1)\pi x}{L(t)}]\left\{3~cosec^{2}(\frac{\pi x}{L(t)})-(n^2+2n+3)\right\}\right.\\ \\
\left.-3(n+1) cot(\frac{\pi x}{L(t)})cos[\frac{(n+1)\pi x}{L(t)}]\right) ~~,~~n\neq 0,1
\ea
\eeq

where $\epsilon_{n2}=\epsilon_{n0} =\epsilon_{n}=\pi^2n(n+2)$.
\vspace{.25cm}

\un{\bf{Construction of $V_{2}(q)$ for $j=1$}}:

For the construction of the potential $V_2(q)$, here we shall require the first and second excited states of the potential $V_0(q)$ and they are given by 
\beq
Q_{1}^{0} = sin(2\pi q)~~and~~~Q_{2}^{0} = sin(3\pi q)\label{q02}\eeq
Thus from (\ref{v2}) we find 
\beq
V_{2}(q) = \frac{\pi^2[135 + 160~cos(2\pi q)+4~cos(4\pi q)+cos(6\pi q)]cosec^{2}(\pi q)}{2(3 + 2~cos(2\pi q))^2}\label{v202}\eeq
and 
\beq
\ba{l}
Q_{n}^{2}(q)=\frac{\pi^2 cosec^{2}(\pi q)}{2(3 + 2~cos(2\pi q))}\left(cos(\pi q)\left\{(n^2-3n+2)sin[(n+4)\pi q]+(n^2+7n+12)sin[(n-2)\pi q] \right\} \right. \\ \\
\left. + 2(n^2+2n-8)sin[(n+1)\pi q]\right)\\ \label{v2qw02}\ea\eeq
The corresponding exactly solvable time dependent potential and wavefunctions are found to be
\beq
V_{2}(x,t)=\frac{\pi^2[135 + 160~cos(\frac{2\pi x}{L(t)}) + 4~cos(\frac{4\pi x}{L(t)}) + cos(\frac{6\pi x}{L(t)})]cosec^{2}(\frac{\pi x}{L(t)})}{2(3 + 2~cos(\frac{2\pi  x}{L(t)}))^2}-\frac{\ddot{L}(t)}{4L(t)}x^2~~,~~ 0\leq x\leq L(t)\label{v202xt}
\eeq
and 
\beq
\ba{l}
\psi_{n}^{2}(x,t)\propto \frac{1}{\sqrt{L(t)}}e^{\frac{i}{4}\frac{\dot{L}(t)}{L(t)}x^2 -i\epsilon_{n2}\tau(t)}\frac{\pi^2 cosec^{2}(\frac{\pi x}{L(t)})}{2(3+2~cos(\frac{2\pi x}{L(t)}))}\left(cos[\frac{\pi x}{L(t)}]\times \right. \\ \\ 
\left. \left\{(n^2-3n+2)sin[\frac{(n+4)\pi x}{L(t)}]+(n^2+7n+12)sin[\frac{(n-2)\pi x}{L(t)}] \right\}+2(n^2+2n-8)sin[\frac{(n+1)\pi x}{L(t)}] \right) 
\ea
\eeq
where $\epsilon_{n2}=\epsilon_{n0} = \epsilon_{n-} = n(n+2)$, $n\neq 1,2$.

\section{Conclusion} After transforming the moving boundary to a fixed one we obtained a number of exactly solvable potentials with moving boundaries using SUSY formalism. In particular we have applied first as well as second order SUSY formalism to the square well potential and obtained a number of exactly solvable potentials with moving right boundary. We belive the present method may be applied to other potentials to generate new exactly solvable potentials with moving boundary.

\ed